\documentclass[a4paper,11pt]{article}
\usepackage{pos}
\usepackage{bbold}
\usepackage{tikz}
\usepackage{float}
\usepackage{booktabs}
\usepackage{multirow}

\newcommand{\dets}{\mathrm{dets}}
\newcommand{\Tr}{\mathrm{Tr}}

\usetikzlibrary{shapes,snakes,arrows,positioning,automata,backgrounds,calc,er,patterns}
\usetikzlibrary{shadings}

\makeatletter
\def\maketag@@@#1{\hbox{\m@th\normalfont\normalsize#1}}
\makeatother

\title{QED and strong isospin corrections in the hadronic vacuum polarization contribution to the anomalous magnetic moment of the muon}
\ShortTitle{QED and strong-isospin-breaking corrections in the LO-HVP contribution to $(g_\mu-2)$}

\author*[f]{L. Parato}
\author[a]{Sz. Borsanyi}
\author[a,b,c,d,e]{Z. Fodor}
\author[a,f]{J. N. Guenther}
\author[a]{C. Hoelbling}
\author[d]{S. D. Katz}
\author[f]{L. Lellouch}
\author[a,b]{T. Lippert}
\author[f,g,h]{K. Miura}
\author[a,b]{K. K. Szabo}
\author[a,b]{F. Stokes}
\author[a]{B. C. Toth}
\author[b]{Cs. Torok}
\author[a,f]{L. Varnhorst}
\author[]{(Budapest-Marseille-Wuppertal collaboration)}

\affiliation[a]{Department of Physics, University of Wuppertal, D-42119 Wuppertal, Germany}
\affiliation[b]{J\"ulich Supercomputing Centre, Forschungszentrum J\"ulich, D-52428 J\"ulich, Germany}
\affiliation[c]{Department of Physics, Pennsylvania State University, University Park, PA 16802, USA}
\affiliation[d]{Institute for Theoretical Physics, E\"otv\"os University, H-1117 Budapest, Hungary}
\affiliation[e]{University of California, San Diego, 9500 Gilman Drive, La Jolla, CA 92093, USA}
\affiliation[f]{Aix Marseille Univ, Universit\'e de Toulon, CNRS, CPT, IPhU, Marseille, France}
\affiliation[g]{Helmholtz Institute Mainz, D-55099 Mainz, Germany}
\affiliation[h]{Kobayashi-Maskawa Institute for the Origin of Particles and the Universe, Nagoya University, Nagoya 464-8602, Japan}

\emailAdd{parato@cpt.univ-mrs.fr}

\abstract{Recently, the Budapest-Marseille-Wuppertal collaboration achieved sub-percent precision 
in the evaluation of the lowest-order hadronic vacuum polarization contribution to the muon $g_\mu-2$ \cite{BMWc2020}. 
At this level of precision, isospin-symmetric QCD is not sufficient. 
In this contribution we review how QED and strong-isospin-breaking effects have been included in our work. 
Isospin breaking is implemented by expanding the relevant correlation functions to second order in the electric charge $e$
and to first order in $m_u-m_d$. 
The correction terms are then computed using isospin-symmetric configurations. 
The choice of this approach allows us to better distribute the available computing resources among the various contributions.}

\FullConference{The 38th International Symposium on Lattice Field Theory, LATTICE2021
26th-30th July, 2021
Zoom/Gather@Massachusetts Institute of Technology}

\begin{document}
\maketitle

\section{Introduction}
The anomalous magnetic moment of the muon, $a_\mu = (g_\mu-2)/2$, is now measured to a precision of 0.35 ppm,
achieved by combining the recent measurement of the Fermilab E989 experiment \cite{E989:2021}
with the previous result of the BNL E821 experiment \cite{E821:2006}:
\begin{equation}
	a_\mu^{\mathrm{Exp}} = 116592061(41) \times 10^{-11}.
\end{equation}
On the theoretical side, the uncertainty on $a_\mu^{\mathrm{SM}}$ is largely dominated by the lowest-order hadronic vacuum polarization 
(LO-HVP) contribution, which accounts for more than $80\%$ of the total theoretical uncertainty and is currently known 
with a relative precision of $0.6\%$ \cite{WhitePaper2020}. 
In order to match the target experimental uncertainty of the Fermilab experiment (0.14 ppm), 
the LO-HVP contribution must be computed with a relative precision of $0.2\%$. 
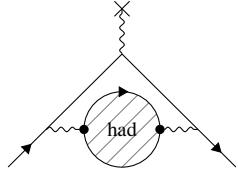
\begin{figure}[H]
	\center
	\begin{tikzpicture}
		[pt/.style={draw=none,fill=none, scale=0.01},
		dot/.style={circle, draw, fill, scale=.3},
		tri0/.style={regular polygon,regular polygon sides=3, draw, fill, color=black, scale=.2,rotate=-90},
		tri1/.style={regular polygon,regular polygon sides=3, draw, fill, color=black, scale=.2,rotate=-45},
		tri2/.style={regular polygon,regular polygon sides=3, draw, fill, color=black, scale=.2,rotate=-135}]
		\node [dot] (a) at (-.5,0) {};
		\node [dot] (b) at (.5,0) {};		
		\draw[color=gray] (a) -- (0,.5);
		\draw[color=gray] (-.45,-.2) -- (.2,.45);
		\draw[color=gray] (-.35,-.35) -- (.35,.35);
		\draw[color=gray] (.45,.2) -- (-.2,-.45);
		\draw[color=gray] (0,-.5) -- (b);
		\node [tri0] (c) at (.0,0.5) {};
		\draw[color=black] (-0,0) circle [radius=.5]; 	
		\node [pt] (a0) at (-1,0) {};
		\node [pt] (b0) at (1,0) {};	
		\tikzset{decoration={snake,amplitude=.3mm,segment length=1.5mm}}
		\draw[decorate,color=black] (a0) -- (a);
		\draw[decorate,color=black] (b) -- (b0);
		\node [pt] (mtop) at (0,1.6) {};
		\node [pt] (m0) at (0,1) {};
		\node [pt] (msx) at (-1.5,-.5) {};
		\node [pt] (mdx) at (1.5,-.5) {};
		\draw[color=black] (msx) -- (m0) -- (mdx);
		\draw[decorate,color=black] (m0) -- (mtop);
		\draw[color=black] (-.1,1.5) -- (.1,1.7);
		\draw[color=black] (-.1,1.7) -- (.1,1.5);
		\node [tri1] (boh) at (-1.25,-.25) {};
		\node [tri2] (boh2) at (1.25,-.25) {};
		\node[text width=1] at (-.175,0) {\scalebox{0.7}{had}};
	\end{tikzpicture}
	\caption{LO-HVP contribution to $a_\mu$}
	\label{fig:hvp}
\end{figure}

A lattice computation at this level of precision cannot be performed using the isospin-symmetric limit of QCD. 
The validity of the $SU(2)_V$ isospin symmetry in QCD relies on the fact that $\delta m/\Lambda_\mathrm{QCD} \equiv (m_d-m_u)/\Lambda_\mathrm{QCD} \sim 0.01$, as well as on the small size of the fine structure constant, $\alpha = e^2/4\pi\sim 1/137$.
Thus, strong-isospin-breaking (SIB) and QED effects become relevant at the percent precision level, and cannot be neglected in a computation aiming at few permil precision.

In our work \cite{BMWc2020}, isospin-breaking (IB) effects are implemented by taking derivatives of QCD+QED expectation values with respect to the bare parameters $e$ and $\delta m/m_{l}$, with $m_l\equiv \tfrac{1}{2}(m_u+m_d)$, and computing the resulting observables on isospin-symmetric configurations, as first proposed in \cite{deDivitiis:2011} and \cite{deDivitiis:2013}.  
The rationale behind this choice is the possibility to optimally distribute the computing resources among the various IB contributions.
IB effects are included in all the observables that enter our analysis: current-current correlators, meson masses needed to fix the physical point, and scale setting. 
The procedure and some details of our calculation are summarized in sections \ref{sec:methodology} and \ref{sec:details}.
Not only do we account for QED and strong isospin-breaking effects in our results, we also perform a separation of isospin symmetric and isospin breaking contributions.
This separation is scheme dependent and requires a convention, which will be outlined in section \ref{sec:IBdecomposition}.  

\section{Methodology}\label{sec:methodology}
Our partition function for 1+1+1+1 staggered fermions is given by
\begin{equation}
	Z = \int [dU]e^{-S_g[U]} \int [dA]e^{-S_\gamma[A]} \, \prod_f det \, M_f[V_U e^{ieq_fA},m_f]^{1/4}
	\label{eq:Z}
\end{equation}
where ${f=\{u,d,s,c\}}$ and $q_f = \left\lbrace \tfrac{2}{3},-\tfrac{1}{3},-\tfrac{1}{3},\tfrac{2}{3}\right\rbrace$.
$U$, $V_U$ and $S_g[U]$ represent the gluon field, the smeared gluon field, and the corresponding gauge action (detailed in Sections 1 and 2 of the \textit{Supplementary Information} of \cite{BMWc2020}).
The photon field $A$ and action $S_\gamma[A]$ are defined in the $\mathrm{QED}_\mathrm{L}$ scheme \cite{Hayakawa2008}.
For convenience, we write the determinant of the fermionic staggered matrix as 
\begin{equation}
	\dets[U,A; \{m_f\}, \{q_f\},e] \equiv \prod_f \det M_f^{1/4}.
\end{equation}
where the explicit form of the fermionic matrix $M_f$ reads
\begin{equation}
	M_f[W,m_f] = \sum_\mu D_\mu[W] + m_f\mathbb{1} \qquad \text{with} \qquad W = V_U e^{ieq_fA}.
\end{equation}

Consider now the observable $X(e,\delta m)$, a function of $e$ and $\delta m$.
We define the derivatives
\begin{equation}
	X_0 = X(0,0), 
	\quad 
	X'_m = m_l \frac{\partial X}{ \partial \delta m} (0,0), 
	\quad 
	X'_1 \equiv \frac{\partial X}{\partial e} (0,0),
	\quad 
	X''_2 \equiv \frac{1}{2}\frac{\partial^2 X}{\partial e^2}(0,0).
\end{equation}
To evaluate the expectation value of $X(e,\delta m)$ to first order in the isospin-breaking parameters $e^2$ and $\delta m$, we expand $X(e,\delta m)$ and the fermionic determinant in $e$ and $\delta m$, including $\mathcal{O}(e^2)$ and $\mathcal{O}(\delta m)$ terms and omitting higher order terms, and evaluate each derivative in the isospin-symmetric limit $e^2 = \delta m = 0$. 
We also make a distinction between the valence charge $e_v$, that appears in the derivatives of the observable $X(e,\delta m)$, and the sea charge $e_s$, that appears in the derivatives of the fermionic determinant.
Explicitly,
\begin{align}
	\left\langle X \right\rangle &\simeq
	\frac{
		\int [dU][dA]e^{-S_g} e^{-S_\gamma} \, \dets_0
		\left(
			\mathbb{1} + e_s\tfrac{\dets^{'}_1}{\dets_0} + e_s^2\tfrac{\dets^{''}_2}{\dets_0} 
		\right)
		\left(
			X_0 + \tfrac{\delta m}{m_l} X'_m +  e_v X^{'}_1 +  e_v^2 X^{''}_2 
		\right) 
	}
	{
		\int [dU][dA]e^{-S_g} e^{-S_\gamma} \, 
		\dets_0  
		\left(
			\mathbb{1} + e_s\tfrac{\dets^{'}_1}{\dets_0} + e_s^2\tfrac{\dets^{''}_2}{\dets_0} 
		\right)
	}
	\\
	&= \langle X_0 \rangle_0 
		+ \tfrac{\delta m}{m_l}\langle X\rangle'_m 
		+ e_v^2 \langle X \rangle''_{20}
		+ e_v e_s \langle X \rangle''_{11}
		+ e^2_s \langle X \rangle''_{02} 
		\label{eq:expansion}
\end{align}
The five terms in \eqref{eq:expansion} correspond to quark-connected (left) and quark-disconnected (right) contractions\footnote{
	Black lines represent quark lines, gluons are implied and not pictured: in particular, disconnected quark loops are to be understood as connected by gluons. 
	The black dots are current insertions, as pictured in Figure \ref{fig:hvp}.
	Blue circles are sea-quark loops generated by the derivatives of $\dets$, yellow lines are photons.
	The red square represents the insertion of the SIB operator. 
}:

\vspace{10pt}\hspace{-16.5pt}
\begin{tabular}{l@{\hskip 40pt}c@{\hskip 40pt}c}
	$\langle X_0\rangle_0 = Z_0^{-1}\int [dU]e^{-S[U]} \dets_0 \,X_0$   
	&
	$\vcenter{\hbox{\begin{tikzpicture}
		[pt/.style={draw=none,fill=none, scale=0.01},
		dot/.style={circle, draw, fill, scale=.3}]
		\node [dot] (a) at (-.35,0) {};
		\node [dot] (b) at (.35,0) {};
		\draw[color=black] (-0,0) circle [radius=.35]; 
		\node [pt] (none1) at (0,-.38) {};
		\node [pt] (none2) at (0,.38) {};		
	\end{tikzpicture}}}$
	&
	$\vcenter{\hbox{\begin{tikzpicture}
		[dot/.style={circle, draw, fill, scale=.3},
		pt/.style={draw=none,fill=none, scale=0.01}]
		\node [dot] (a) at (-.55,0) {};
		\node [dot] (b) at (.55,0) {};
		\draw[color=black] (-.3,0) circle [radius=.25]; 
		\draw[color=black] (.3,0) circle [radius=.25]; 
		\node [pt] (none1) at (0,-.38) {};
		\node [pt] (none2) at (0,.38) {};							
	\end{tikzpicture}}}$
	\\ [15pt]
	$\langle X \rangle'_{m} \equiv \left\langle X'_m \right\rangle_0$
	&
	${\vcenter{\hbox{\begin{tikzpicture}
		[sq/.style={regular polygon,regular polygon sides=4, draw, fill, color=purple, scale=.3},
		pt/.style={draw=none,fill=none, scale=0.01},
		dot/.style={circle, draw, fill, scale=.3}]
		\node [dot] (b) at (-.35,0) {};
		\node [dot] (c) at (.35,0) {};
		\draw[color=black] (-0,0) circle [radius=.35]; 
		\node [sq] (a) at (0,0.35) {};
		\node [pt] (none1) at (0,-.38) {};
		\node [pt] (none2) at (0,.38) {};		
	\end{tikzpicture}}}}$
	&
	${\vcenter{\hbox{\begin{tikzpicture}
		[sq/.style={regular polygon,regular polygon sides=4, draw, fill, color=purple, scale=.3},
		pt/.style={draw=none,fill=none, scale=0.01},
		dot/.style={circle, draw, fill, scale=.3}]
		
		\node [dot] (b) at (-.55,0) {};
		\node [dot] (c) at (.55,0) {};
		\draw[color=black] (-.3,0) circle [radius=.25]; 
		\draw[color=black] (.3,0) circle [radius=.25]; 
		\node [sq] (a) at (-.3,.25) {};
		
		\node [pt] (none1) at (0,-.38) {};
		\node [pt] (none2) at (0,.38) {};					
	\end{tikzpicture}}}}$
	\\ [15pt]
	$\langle X \rangle''_{20} \equiv \left\langle X''_{2} \right\rangle_0$
	&
	${\vcenter{\hbox{\begin{tikzpicture}
		[pt/.style={draw=none,fill=none, scale=0.01},
		dot/.style={circle, draw, fill, scale=.3}]
		\node [pt] (a) at (0,0.35) {};
		\node [pt] (b) at (0,-0.35) {};
		\node [dot] (c) at (-.35,0) {};
		\node [dot] (d) at (.35,0) {};
		\tikzset{decoration={snake,amplitude=.5mm,segment length=1.5mm}}
		\draw[decorate,color={rgb,255:red,222; green,175; blue,0}] (a) -- (b);
		\draw[color=black] (-0,0) circle [radius=.35]; 
	\end{tikzpicture}}}}
	\hspace*{20pt}
	{\vcenter{\hbox{\begin{tikzpicture}
		[pt/.style={draw=none,fill=none, scale=0.01},
		dot/.style={circle, draw, fill, scale=.3}]
		\node [pt] (a) at (0.3,0.2) {};
		\node [pt] (b) at (-0.3,0.2) {};
		\node [dot] (c) at (-.35,0) {};
		\node [dot] (d) at (.35,0) {};
		\tikzset{decoration={snake,amplitude=.5mm,segment length=1.5mm}}
		\draw[decorate,color={rgb,255:red,222; green,175; blue,0}] (a) -- (b);
		\draw[color=black] (-0,0) circle [radius=.35]; 
	\end{tikzpicture}}}}$
	&
	${
	\vcenter{\hbox{\begin{tikzpicture}
		[pt/.style={draw=none,fill=none, scale=0.01},
		dot/.style={circle, draw, fill, scale=.3}]
		\node [pt] (a) at (-.3,.25) {};
		\node [pt] (d) at (-.3,-.25) {};
		\node [dot] (b) at (-.55,0) {};
		\node [dot] (c) at (.55,0) {};
		\tikzset{decoration={snake,amplitude=.5mm,segment length=1.5mm}}
		\draw[decorate,color={rgb,255:red,222; green,175; blue,0}] (a) -- (d);
		\draw[color=black] (-.3,0) circle [radius=.25]; 
		\draw[color=black] (.3,0) circle [radius=.25]; 
		\node [pt] (none1) at (0,-.35) {};
		\node [pt] (none2) at (0,.35) {};
	\end{tikzpicture}}}}
	\hspace*{20pt}
	{\vcenter{\hbox{\begin{tikzpicture}
		[pt/.style={draw=none,fill=none, scale=0.01},
		dot/.style={circle, draw, fill, scale=.3}]
		\node [pt] (a) at (-.15,0) {};
		\node [pt] (d) at (.15,0) {};
		\node [dot] (b) at (-.65,0) {};
		\node [dot] (c) at (.65,0) {};
		\tikzset{decoration={snake,amplitude=.5mm,segment length=1.8mm}}
		\draw[decorate,color={rgb,255:red,222; green,175; blue,0}] (a) -- (d);
		\draw[color=black] (-.4,0) circle [radius=.25]; 
		\draw[color=black] (.4,0) circle [radius=.25]; 
		\node [pt] (none1) at (0,-.35) {};
		\node [pt] (none2) at (0,.35) {};
	\end{tikzpicture}}}}
	$			
	\\ [15pt]
	$\langle X \rangle''_{11} \equiv \left\langle X'_{1} \tfrac{\dets'_1}{\dets_0} \right\rangle_0$
	&
	${\vcenter{\hbox{\begin{tikzpicture}
		[pt/.style={draw=none,fill=none, scale=0.01},
		dot/.style={circle, draw, fill, scale=.3}]
		\node [pt] (c) at (0,.35) {};
		\node [pt] (d) at (0,.5) {};
		\node [dot] (a) at (-.35,0) {};
		\node [dot] (b) at (.35,0) {};
		\tikzset{decoration={snake,amplitude=.5mm,segment length=.8mm}}
		\draw[decorate,color={rgb,255:red,222; green,175; blue,0}] (c) -- (d);
		\draw[color={rgb,100:red,14; green,68; blue,92}] (0,0.7) circle [radius=.2]; 
		\draw[color=black] (-0,0) circle [radius=.35]; 
	\end{tikzpicture}}}}$
	&
	${\vcenter{\hbox{\begin{tikzpicture}
		[pt/.style={draw=none,fill=none, scale=0.01},
		dot/.style={circle, draw, fill, scale=.3}]
		\node [dot] (b) at (-.55,0) {};
		\node [dot] (c) at (.55,0) {};
		\node [pt] (a) at (-0.3,.25) {};
		\node [pt] (d) at (-0.2,.5) {};
		\draw[color=black] (-.3,0) circle [radius=.25]; 
		\draw[color=black] (.3,0) circle [radius=.25]; 
		\tikzset{decoration={snake,amplitude=.5mm,segment length=1.5mm}}
		\draw[decorate,color={rgb,255:red,222; green,175; blue,0}] (a) -- (d);
		\draw[color={rgb,100:red,14; green,68; blue,92}] (0,0.5) circle [radius=.2]; 
	\end{tikzpicture}}}}$	
	\\ [15pt]
	$\langle X \rangle''_{02} \equiv \left\langle \left[ X_0 - \langle X_0 \rangle_0\right] \tfrac{\dets''_2}{\dets_0}\right\rangle_0$
	&
	${\vcenter{\hbox{\begin{tikzpicture}
		[pt/.style={draw=none,fill=none, scale=0.01},
		dot/.style={circle, draw, fill, scale=.3}]
		\node [dot] at (-.35,0) {};
		\node [dot] at (.35,0) {};
		\node [pt] (a) at (0.2,.6) {};
		\node [pt] (b) at (-0.2,.6) {};
		\draw[color=black] (-0,0) circle [radius=.35]; 
		\draw[color={rgb,100:red,14; green,68; blue,92}] (0,.6) circle [radius=.2]; 
		\tikzset{decoration={snake,amplitude=.5mm,segment length=1.5mm}}
		\draw[decorate,color={rgb,255:red,222; green,175; blue,0}] (a) -- (b);
	\end{tikzpicture}}}}
	\hspace*{20pt}
	{\vcenter{\hbox{\begin{tikzpicture}
		[pt/.style={draw=none,fill=none, scale=0.01},
		dot/.style={circle, draw, fill, scale=.3}]
		\node [dot] at (-.35,0) {};
		\node [dot] at (.35,0) {};
		\node [pt] (a) at (0.15,.6) {};
		\node [pt] (b) at (-0.15,.6) {};
		\draw[color=black] (-0,0) circle [radius=.35]; 
		\draw[color={rgb,100:red,14; green,68; blue,92}] (-0.35,.6) circle [radius=.2]; 
		\draw[color={rgb,100:red,14; green,68; blue,92}] (0.35,.6) circle [radius=.2]; 
		\tikzset{decoration={snake,amplitude=.5mm,segment length=1.25mm}}
		\draw[decorate,color={rgb,255:red,222; green,175; blue,0}] (a) -- (b);
	\end{tikzpicture}}}}$				
	&
	${\vcenter{\hbox{\begin{tikzpicture}
		[pt/.style={draw=none,fill=none, scale=0.01},
		dot/.style={circle, draw, fill, scale=.3}]
		\node [dot] (d) at (-.55,0) {};
		\node [dot] (c) at (.55,0) {};
		\draw[color=black] (-.3,0) circle [radius=.25]; 
		\draw[color=black] (.3,0) circle [radius=.25]; 
		\node [pt] (a) at (0.2,.55) {};
		\node [pt] (b) at (-0.2,.55) {};
		\draw[color={rgb,100:red,14; green,68; blue,92}] (0,.55) circle [radius=.2]; 
		\tikzset{decoration={snake,amplitude=.5mm,segment length=1.5mm}}
		\draw[decorate,color={rgb,255:red,222; green,175; blue,0}] (a) -- (b);
	\end{tikzpicture}}}}
	\hspace*{20pt}
	{\vcenter{\hbox{\begin{tikzpicture}
		[pt/.style={draw=none,fill=none, scale=0.01},
		dot/.style={circle, draw, fill, scale=.3}]
		\node [dot] (d) at (-.55,0) {};
		\node [dot] (c) at (.55,0) {};
		\draw[color=black] (-.3,0) circle [radius=.25]; 
		\draw[color=black] (.3,0) circle [radius=.25]; 
		\node [pt] (a) at (0.15,.55) {};
		\node [pt] (b) at (-0.15,.55) {};
		\draw[color={rgb,100:red,14; green,68; blue,92}] (-0.35,.55) circle [radius=.2]; 
		\draw[color={rgb,100:red,14; green,68; blue,92}] (0.35,.55) circle [radius=.2]; 
		\tikzset{decoration={snake,amplitude=.5mm,segment length=1.25mm}}
		\draw[decorate,color={rgb,255:red,222; green,175; blue,0}] (a) -- (b);
	\end{tikzpicture}}}}$								
\end{tabular}
\vspace{10pt}

Note that \eqref{eq:expansion} is an expansion in bare parameters and not an IB decomposition of $\langle X\rangle$.
The latter requires us to define a suitable set of physical observables to separate the contributions,  which will be discussed in section \ref{sec:IBdecomposition}. 

Note also that $\dets'_m=0$, which comes from the symmetry of $\dets$ under the exchange of $m_d=m_l+\delta m/2$ and $m_u=m_l-\delta m/2$ at $e_s=0$.
The first and second derivatives of $\dets$ with respect to $e_s$ are given by:
\begin{small}
\begin{align}
	\frac{\dets'_1}{\dets_0} &= \sum_f \frac{q_f}{4} \Tr\left\lbrace M_f^{-1}D[iAV_U] \right\rbrace
	\\	
	\frac{\dets''_2}{\dets_0} &=
	\frac{1}{2} 
	\left[ 
	\left( \frac{\dets'_1}{\dets_0} \right)^2
	-\sum_f \frac{q_f^2}{4} \Tr  \left\lbrace M_f^{-1} D[iAV_U]M_f^{-1} D[iAV_U] \right\rbrace
	-\sum_f \frac{q_f^2}{4} \Tr \left\lbrace M_f^{-1} D[A^2V_U] \right\rbrace 
	\right]
\end{align}
\end{small}\\
Diagrammatically,
$
	\dets'_1/\dets_0 = \vcenter{\hbox{
		\begin{tikzpicture}
		[pt/.style={draw=none,fill=none, scale=0.001},
		cross/.style={cross out, draw,color={rgb,255:red,222; green,175; blue,0}, scale=.5}]
		\node [cross] (a) at (0.17,0) {};
		\node [pt] (a) at (0.15,0) {};
		\node [pt] (b) at (-0.15,0) {};
		\draw[color={rgb,100:red,14; green,68; blue,92}] (-0.35,0) circle [radius=.2];  
		\tikzset{decoration={snake,amplitude=.5mm,segment length=1.5mm}}
		\draw[decorate,color={rgb,255:red,222; green,175; blue,0}] (a) -- (b);
		\end{tikzpicture}
		}}
$,
hence
$
	\left\langle \dets''_2/\dets_0 \right\rangle = 
	\frac{1}{2}\left[
	\vcenter{\hbox{\begin{tikzpicture}
		[pt/.style={draw=none,fill=none, scale=0.001},
		cross/.style={cross out, draw,color={rgb,255:red,222; green,175; blue,0}, scale=.3}]
		\node [pt] (a) at (0.15,0) {};
		\node [pt] (b) at (-0.15,0) {};
		\draw[color={rgb,100:red,14; green,68; blue,92}] (-0.35,0) circle [radius=.2];  
		\draw[color={rgb,100:red,14; green,68; blue,92}] (0.35,0) circle [radius=.2];  
		\tikzset{decoration={snake,amplitude=.5mm,segment length=1.5mm}}
		\draw[decorate,color={rgb,255:red,222; green,175; blue,0}] (a) -- (b);
	\end{tikzpicture}}}
	\,+\,
	\vcenter{\hbox{\begin{tikzpicture}
		[pt/.style={draw=none,fill=none, scale=0.001}]
		\node [pt] (a) at (0.2,0) {};
		\node [pt] (c) at (-0.2,0) {};
		\draw[color={rgb,100:red,14; green,68; blue,92}] (0,0) circle [radius=.2]; 
		\tikzset{decoration={snake,amplitude=.5mm,segment length=1.5mm}}
		\draw[decorate,color={rgb,255:red,222; green,175; blue,0}] (a) -- (c);
	\end{tikzpicture}}}
	\,+\,
	\vcenter{\hbox{\begin{tikzpicture}
		[pt/.style={draw=none,fill=none, scale=0.001}]
		\tikzset{decoration={snake,amplitude=.3mm,segment length=1.54mm}};
		\draw[color={rgb,100:red,14; green,68; blue,92}] (0.37,0) circle [radius=.2]; 
		\draw[decorate, color={rgb,255:red,222; green,175; blue,0},rotate=80] (0,0) circle [radius=.175]; 
	\end{tikzpicture}}}
	\right]
$.
In the above calculations we have used 
$
	\partial_{e_s}\det M_f 
	= \partial_{e_s} \exp\left(\Tr\lbrace\ln M_f\rbrace\right)  
	= \det M_f ~ \Tr \lbrace M_f^{-1}\partial_{e_s}\,M_f \rbrace
$ 
and $\partial_{e_s} M_f = q_f D_\mu[iA V_U \exp(ieq_fA)]$. 
We refer to all $e_s$-dependent terms in the expansion of an observable as \emph{dynamical QED contributions}.
We use random sources, a truncated solver method \cite{Bali2010, Blum2013}, and low-mode averaging \cite{Neff2001,Li2010} to efficiently compute {$\dets'_1$} and {$\dets''_2$}. 

\section{Computation of isospin-breaking derivatives}\label{sec:details}
In this section we review how the isospin-breaking derivatives of the hadron masses and of the current propagator have been computed in our work.
Before going into the detail of each derivative, let us introduce some useful notations and observations.  

\paragraph{Current propagator.} 
Given the generating functional $Z[A^{\mathrm{ext}}]=\int ... \dets[U,A+A^{\mathrm{ext}},\{q_f\},\{m_f\},e]$, the conserved current propagator can be computed as
\begin{equation}
	\langle J_{\mu x} J_{\bar{\mu}\bar{x}}\rangle/e^2 
	\equiv \frac{1}{e^2} \frac{\delta^2 \log Z}{\delta A^{\mathrm{ext}}_{\mu,x} A^{\mathrm{ext}}_{\bar{x}\bar{\mu}}} \bigg\vert_{A^{\mathrm{ext}}=0}
	=
	\left\langle 
	\sum_f q_f^2 C^{\mathrm{conn}}_{\mu,x,\bar{\mu},\bar{x}}(m_f, eq_f) 
	+
	C^{\mathrm{disc}}_{\mu,x,\bar{\mu},\bar{x}}
	+
	c.t.  
	\right\rangle,
\end{equation}
where $c.t.$ is a contact term that does not contribute to the observables of interest. 
The explicit form of the other two terms is given by
\begin{equation}
	C^{\mathrm{conn}}_{\mu,x,\bar{\mu},\bar{x}}(m_f, eq_f) \equiv -\frac{1}{4} 
	\Tr \left\lbrace M_f^{-1} D_\mu[iP_xV_U e^{ieq_fA}] M_f^{-1} D_{\bar{\mu}}  [iP_{\bar{x}} V_U e^{ieq_fA}] \right\rbrace
	\label{eq:conn}
\end{equation}
for the connected contraction, and 
\begin{equation}
	C^\mathrm{disc}_{\mu,x,\bar{\mu},\bar{x}} \equiv 
	\sum_{f,\bar{f}} q_f q_{\bar{f}} 	I_{\mu,x}(m_f,eq_f) I_{\bar{\mu},\bar{x}} (m_{\bar{f}},eq_{\bar{f}})
	\label{eq:disc}
\end{equation}
for the disconnected contraction, where 
\begin{equation}
	I_{\mu,x}(m_f,eq_f) \equiv \frac{1}{4} \Tr \left\lbrace M_f^{-1} D_\mu[iP_xV_U e^{ieq_fA}]\right\rbrace.
	\label{eq:single_disc}
\end{equation}
In the above calculations we have used $\delta\det M_f/\delta A_{\mu,x} = \det M_f  ~
\Tr \lbrace M_f^{-1} (\delta\,M_f/\delta\,A_{\mu,x}) \rbrace$ 
and $\delta\,M_f/\delta\,A_{\mu,x} = D_\mu[iP_xV_U \exp(ieq_fA)]$, where $P_x$ is the projection operator, which sets to zero all components of the argument of $D_\mu$ which are not at $x$. 
We further split the connected part in 
\begin{align}
	C^{\mathrm{light}} &\equiv \tfrac{4}{9} C^{\mathrm{conn}} (m_u, \tfrac{2}{3}e) + \tfrac{1}{9}C^{\mathrm{conn}}(m_d,-\tfrac{1}{3}e),
	\\
	C^{\mathrm{strange}} &\equiv \tfrac{1}{9} C^{\mathrm{conn}} (m_s,-\tfrac{1}{3}e),
	\\
	C^{\mathrm{charm}} &\equiv \tfrac{4}{9} C^{\mathrm{conn}} (m_c, \tfrac{2}{3}e).
\end{align}
where we drop, for brevity, the Lorentz indices and the coordinates.
With the same omission of subscripts, we shorten \eqref{eq:disc} as $C^\mathrm{disc}$.

Note: when discussing isospin-breaking corrections to the current propagator, we consider the contribution of the light and strange quarks only.
Leading-order electromagnetic corrections to $C^{\mathrm{charm}}$ were computed in \cite{Giusti2017},  whereas the effects of valence charm quarks on $C^\mathrm{disc}$ were estimated in \cite{Borsanyi2017}: on the coarsest lattice, they affect the result by a value much smaller than the statistical error.

\paragraph{Hadron masses.}
We denote a hadron mass by $M = \mathcal{M}[\langle H \rangle]$, where $\mathcal{M}$ is the function needed to extract the mass from the hadron propagator $H$. 
$\mathcal{M}$ is chosen such that the derivatives $\tfrac{\delta M}{\delta H}$ can be given in closed analytic form.

\begin{figure}
	\begin{center}
		\includegraphics[scale=.7]{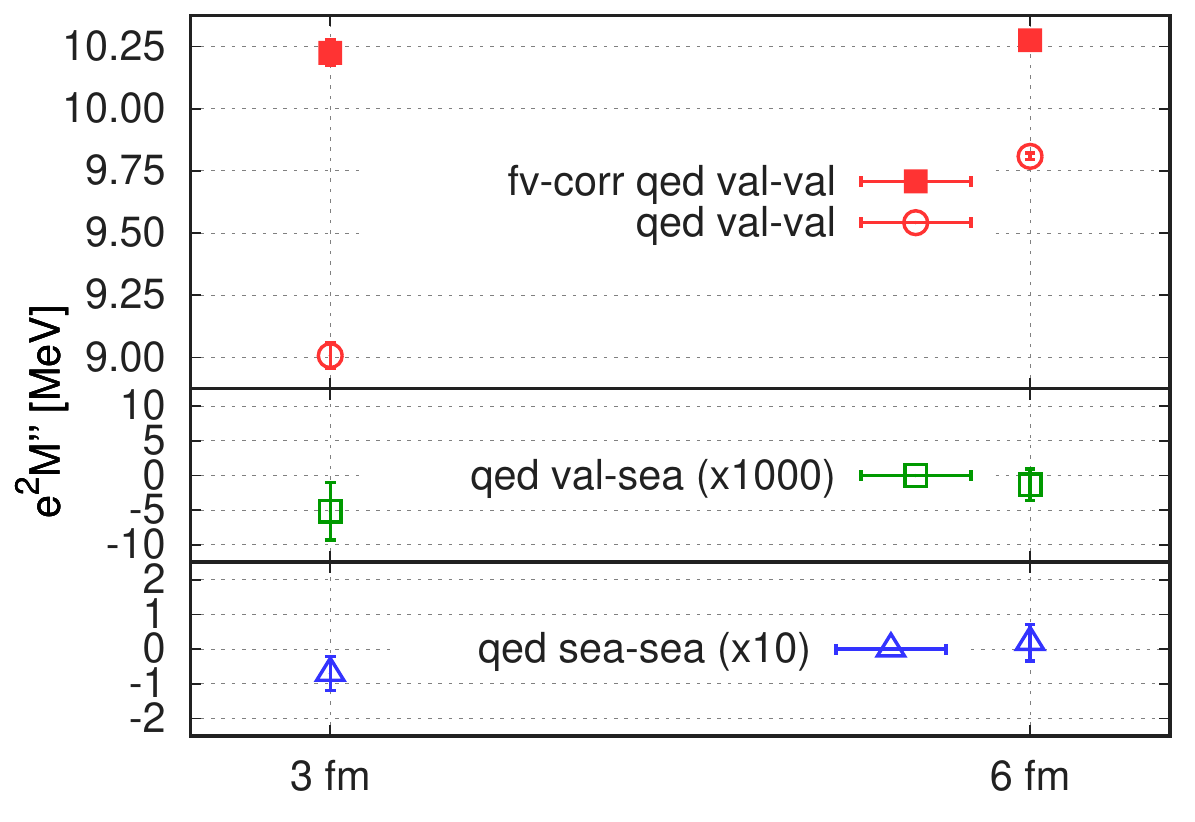}
	\end{center}
	\caption{Volume dependence of QED contributions to the $\pi^+$ mass at $\beta=3.7000$. The red squares in the upper pane correspond to $M''_{20}$ corrected with \eqref{eq:QEDLvol}. The valence-sea $M''_{11}$ and sea-sea $M''_{02}$ contributions are multiplied by 1000 and 10 on the plot, respectively.}
	\label{fig:FVE}
\end{figure}

\paragraph{$\mathbf{QED_L}$ volume effects.}
In the computation of QED derivatives, hadron masses are affected by $\mathcal{O}(1/L)$ volume effects, due to the $\textrm{QED}_L$ scheme \cite{Davoudi2014,Borsanyi2015}. 
The first two orders in $1/L$ are known analytically and depend only on the mass $M$ and electric charge $Q$ of the hadron:
\begin{equation}
	M(L) - M  = -\frac{(Qe)^2c}{8\pi} \left[ \frac{1}{L} + \frac{2}{ML^2} + O(L^{-3})\right] 
	\quad \mathrm{with} \quad
	c=2.837297...
	\label{eq:QEDLvol}
\end{equation}
Valence-valence QED effects (as well as SIB effects) are evaluated on a subset of the set of ensembles of size $L=6$ fm used for the main part of the computation, the isospin-symmetric one.
Dynamical QED effects are evaluated on a dedicated set of ensembles of size $L=3$ fm.
Measurements on $L=3$ fm boxes require one order magnitude less computer time than ones on $L=6$ fm boxes,  for the same level of precision. 
On our coarsest lattice, at $\beta=3.700$, all QED contributions are computed in both volumes.
Dynamical contributions $M''_{11}$ and $M''_{02}$ do not show statistically relevant differences between  results obtained in $L=3$ fm and $L=6$ fm boxes.
Valence-valence derivatives $M''_{20}$, instead, show significant volume dependence (see Figure \ref{fig:FVE}).
$M''_{20}$ terms are thus corrected using the first two orders of \eqref{eq:QEDLvol}.

\subsection{Strong isospin-breaking derivatives\texorpdfstring{$\mathbf{: \langle X \rangle'_m}$}{}}
\begin{itemize}
	\item $[C^{\text{light}}]'_m$ is computed via insertion of the operator corresponding to the mass derivative.
	Since the light propagator is noisy, we evaluate $[C^{\text{conn}}(\kappa m_l,0)]'_m$ at multiples $\kappa$ of the light-mass, with $\kappa = 3,5,7,9,11$. 
	Then, we perform a chiral extrapolation to $\kappa = 1$ to get $[C^{\text{light}}]'_m$.
	\item To evaluate the SIB derivative of the disconnected current propagator, we first observe that 
	\begin{equation}
		[C^{\mathrm{disc}}]'_m \equiv m_l 
		\frac{\partial C^\mathrm{disc}}{\partial\delta m} \bigg\vert_{\delta m = e = 0} = 
		-\frac{3}{2} m_l \frac{\partial C^\mathrm{disc}_0}{\partial m_l}.
	\end{equation}
	The derivative $\partial C^\mathrm{disc}_0/\partial m_l$ is computed as a finite difference: it is sufficient to evaluate, another light-trace with a slightly different $m_l$. We use $I(m_l,0)$ and $I(0.9~m_l,0)$.
	\item The SIB derivative of hadron masses is computed as the finite difference 
	\begin{equation}
		\hspace{20pt}
		M'_m \approx \frac{m_l}{\delta m}  \left( \mathcal{M}[\langle H_{\delta m}\rangle_0] - \mathcal{M}[\langle H_0\rangle_0] \right)		
	\end{equation}
	where $H_{\delta m}$ is the hadron propagator evaluated at $\delta m=2m_l\tfrac{1-r}{1+r}$ with $r=\tfrac{m_u}{m_d}=0.485$ \cite{Fodor:2016bgu} and $e_v=0$.
\end{itemize}

\subsection{Electromagnetic valence-valence derivatives\texorpdfstring{: $\mathbf{\langle X \rangle''_{20}}$}{}}
\begin{itemize}
	\item The second derivative $X''_2$ of an observable $X(e,\delta m)$
	can be computed as the finite difference $X''_2 \approx \tfrac{1}{2e_v}\left[(X_++X_-)-2X_0\right]$, where $X_\pm \equiv X(\pm e_*,0)$ and $e_*$ is the physical value of the electric coupling.
	\item In the valence-valence case, QED is inserted as a stochastic $U(1)$ field \cite{Duncan1996}.
	The statistical noise contributes at order $\mathcal{O}(e)$, but it can be removed by averaging, for each gauge configuration, the two evaluations performed at $\pm e_v$ \cite{Blum2007}.
	\item 
	To evaluate $[C^{\text{strange}}]''_{20}$, we compute $C^{\text{conn}}(m_s,0)$ and $C^{\text{conn}}(m_s,\pm \tfrac{1}{3}e_*)$.
	\item To evaluate $[C^{\text{light}}]''_{20}$, we compute $C^{\text{conn}}(\kappa m_l,0)$ and $C^{\text{conn}}(\kappa m_l,\pm \tfrac{1}{3}e_*)$ for $\kappa = 3,5,7,9,11$,	then we perform a chiral extrapolation to $\kappa = 1$, as for $[C^{\text{light}}]'_m$.
	\item The valence-valence derivative of the current disconnected contraction \eqref{eq:disc} can be computed by rewriting the single contraction,
	$
		\sum_f q_fI_{\mu,x}(m_f,eq_f) = \tfrac{2}{3}I(m_l,\tfrac{2}{3}e)-\tfrac{1}{3}I(m_l,-\tfrac{1}{3}e)-\tfrac{1}{3}I(m_s,\tfrac{1}{3}e)
	$, 
	as 
	$
		-2 I(m_l,0)+2 I(m_l,\tfrac{1}{3}e)+\tfrac{1}{3}I(m_l,-\tfrac{1}{3}e)-\tfrac{1}{3}I(m_s,\tfrac{1}{3}e)
	$, 
	which equals the first expression when expanding in $e$ up to and including $\mathcal{O}(e^2)$ terms. 
	The computation of 
	$
		I(m_l,0)$, $I(m_l,\pm\tfrac{1}{3}e_*)$, $I(m_s,0)
	$, and 
	$
		I(m_s,\pm\tfrac{1}{3}e_*)
	$ 
	is thus sufficient to compute $[C^{\text{disc}}]''_{20}$ as a linear combination of finite differences. 
	\item Valence-valence derivatives of hadron masses are computed as the finite difference 
	{\begin{small}
	\begin{equation}
		M_{20}'' \approx 
		\frac{1}{2e_v^2}
		\left(
			\mathcal{M}[\langle H_+\rangle_0] + \mathcal{M}[\langle H_-\rangle_0] 
			-2 \mathcal{M}[\langle H_0\rangle_0]
		\right)
		=
		\frac{1}{e_v^2}
		\left(
			\mathcal{M}[\tfrac{1}{2}\langle H_+ + H_-\rangle_0] - 
			\mathcal{M}[\langle H_0\rangle_0]
		\right)
	\end{equation}
	\end{small}}
	\vspace{-10pt}
	\\
	and corrected for the finite-volume effects induced by the $\mathrm{QED}_L$ scheme, as mentioned above.
	Here $H_\pm$ is the hadron propagator measured at $\delta m=0$ and $e_v=\pm e_*$.
\end{itemize}

\subsection{Electromagnetic sea-valence derivatives\texorpdfstring{: $\mathbf{\langle X \rangle''_{11}}$}{}}
\begin{itemize}
	\item Sea-valence contributions are evaluated as
	\begin{equation}
		\langle X \rangle''_{11} = \left\langle \left\langle X'_1 
		\frac{\dets'_1}{\dets_0}\right\rangle_A \right\rangle_U,
	\end{equation}
	where the subscript $A$ indicates an average over free photon fields, sampled from $e^{-S_\gamma}$, and the subscript $U$ an average over dynamical gluon configurations.	
	To estimate the first derivative $\dets'_1/\dets_0$,
	we generate one photon field $A$ for each gluon field $U$ 
	and $\sim 10^4$ random vectors on each $(U, A)$ pair.
	$X'_1$ is evaluated as a finite difference: $X'_1\approx \tfrac{1}{2e_v}(X_+ - X_-)$.
	\item Hadron masses are given in the mixed form
	\begin{equation}
		M''_{11} = \frac{\delta \mathcal{M}}{\delta H}\bigg\vert_{\langle H_+ + H_-\rangle_0} \cdot \left\langle \frac{H_+ - H_-}{2e_v} \frac{\dets'_1}{\dets_0}\right\rangle,
	\end{equation}
	with $H_{\pm}$ the hadron propagator evaluated at $\pm e_*$ and $\delta m=0$.
\end{itemize}

\subsection{Electromagnetic sea-sea derivatives\texorpdfstring{: $\mathbf{\langle X \rangle''_{02}}$}{}}

\begin{itemize}
	\item Sea-sea contributions are evaluated as
	\begin{equation}
		\langle X \rangle''_{02} = \left\langle  [X_0 - \langle X_0\rangle_U]\left\langle
		\frac{\dets''_2}{\dets_0}\right\rangle_A \right\rangle_U.
	\end{equation}
	To estimate the second derivative $\dets''_2/\dets_0$,
	we generate $\sim 2000$ photon fields for each gluon field $U$,
	and 12 random sources for each photon field $A$.
	\item Hadron masses are again given in a mixed form:
	\[{M''_{02} = \frac{\delta\mathcal{M}}{\delta H}\bigg\vert_{\langle H_0 \rangle_0} \cdot \left\langle (H_0 - \langle H_0\rangle) \frac{\dets''_2}{\dets_0} \right\rangle_0}. \]
\end{itemize}

\section{Isospin-breaking decomposition}\label{sec:IBdecomposition}

\paragraph{Type-I fits (global fits)}
Our simulation has five bare QCD parameters that need to be fixed: $\{a,m_l,\delta m,m_s,e\}$, plus the charm mass which is fixed by the strange mass via $m_c \equiv 11.85 \,m_s$.

Given the following definitions for lattice observables (left) and their physical values (right), that can be obtained from the PDG \cite{PDG2020},
\begin{equation}
	\begin{cases}
		M_{\pi_\chi}^2  &\equiv \tfrac{1}{2}(M_{uu}^2 + M_{dd}^2)\\
		M^2_{K_\chi} &\equiv \tfrac{1}{2} \left( M_{ds}^2 + M_{us}^2 - M_{ud}^2 \right)\\
		\Delta M_K^2 &\equiv M_{ds}^2 - M_{us}^2
	\end{cases}
	\qquad 
	\begin{cases}
		[M^2_{\pi_\chi}]_* &\approx [M^2_{\pi_0}]_* \\
		[M^2_{K_\chi}]_* &\equiv \tfrac{1}{2}\left([M^2_{K_0}]_* + [M^2_{K_+}]_* - [M^2_{\pi_+}]_*\right) \\
		[\Delta M^2_K]_* &\equiv [M^2_{K_0}]_* - [M^2_{K_+}]_*
	\end{cases}
	\label{eq:scattering_masses}
\end{equation}
(with mesons $M_{uu}$ and $M_{dd}$ defined below),
a possible way of interpolating to the physical point is to set $a$ using the $\Omega^{-}$ baryon mass $M_{\Omega^-}$, the electric charges to $e_v \equiv e_s \equiv e_* = \sqrt{4\pi\alpha_*}$ 
(with $ \alpha_*$ the experimental value of the fine structure constant, a choice that is valid at leading order in isospin-breaking), and to fix the quark masses by studying the dependence of the observables of interest on $M_{\pi_\chi}^2$, $M^2_{K_\chi}$ and $\Delta M_K^2$ around their physical values:
\begin{align}
	\begin{cases}
	a = (a M_{\Omega^-})/[M_{\Omega^-}]_* 
	\\
	(X_{vv} \equiv e_v^2) = (X_{vs} \equiv e_ve_s) = (X_{ss} \equiv e_s^2) = 4\pi\alpha_*
	\\
	m_l \mid  X_l \equiv \frac{M_{\pi_\chi}^2}{M_\Omega^2} - \frac{[M_{\pi_\chi}^2]_*}{{M_\Omega^2}_*} \quad\text{scatters around}\quad 0
	\\
	m_s \mid X_s \equiv \frac{M_{K_\chi}^2}{M_\Omega^2} - \frac{[M_{K_\chi}^2]_*}{{M_\Omega^2}_*} \quad\text{scatters around}\quad 0
	\\
	\delta m \mid X_{\delta m} \equiv \frac{\Delta M^2_K}{M_\Omega^2}  \quad\text{scatters around}\quad \frac{[\Delta M^2_K]_*}{[M_\Omega^2]_*}
	\end{cases}
\end{align}
The specific values of $m_l, m_s$ and $\delta m$ are chosen to be slightly different on each ensemble,  such that, altogether, the various measurements bracket the physical point.  
Thus, one can proceed by parametrizing an observable of interest $O$ with a linear function:
\begin{equation}
	O = f(\{X\},A,B,...) \equiv A + BX_l + CX_s + DX_{\delta m} + EX_{vv} + FX_{vs} + GX_{ss},
	\label{eq:lineartype1}
\end{equation}
where the fit coefficients $A, B, C, F, G$ are polynomials in $a^2$, while $D$ and $E$ can also depend on $X_l$ and $X_s$.
For example, $A = A_0 + A_2a^2 + A_4a^4$.
The continuum extrapolation is then given by 
\begin{equation}
	O_* = A_0 + D_0[X_{\delta m}]_* + (E_0 + F_0 + G_0)e_*^2.
\end{equation}

This kind of parametrization, with experimentally measurable quantities as input, is referred as \emph{type-I} (see Section 3 of \cite{Varnhorst2021LAT} or Section 20 of \cite{BMWc2020} for more details). 
It is not suitable for an isospin-breaking decomposition (note for example that the observable most sensitive to strong-isospin-breaking effects, $\Delta M_K^2$, is also charged). 

\paragraph{Type-II fits (decomposition-friendly parametrization).}
We introduce a second parametrization in order to obtain the decomposition of observables into isospin-symmetric and isospin-breaking parts. 
We fix the bare parameters $\{a,m_l,\delta m,m_s,e\}$ by a new set of observables that we impose to be equal 
in the isospin-symmetric and full QED+QCD theory:
\begin{align}
	\begin{cases}
	a = [w_0]_*/(w_0/a)
	\\
	(\tilde{X}_{vv} \equiv e_v^2) = (\tilde{X}_{vs} \equiv e_ve_s) = (\tilde{X}_{ss} \equiv e_s^2) = 4\pi\alpha_*
	\\
	m_l \mid  \tilde{X}_l \equiv M_{\pi_\chi}^2 w_0^2 - [M_{\pi_\chi}^2w_0^2]_* \quad\text{scatters around}\quad 0
	\\
	m_s \mid \tilde{X}_s \equiv M_{ss}^2 w_0^2- [M_{ss}^2 w_0^2]_* \quad\text{scatters around}\quad 0
	\\
	\delta m \mid \tilde{X}_{\delta m} \equiv \Delta M^2w_0^2 \quad\text{scatters around}\quad [\Delta M^2w_0^2]_* 
	\end{cases}
\end{align}
where 
\begin{itemize}
	\item $w_0$ is the Wilson-flow–based, pure-gauge scale defined in \cite{BMW2012:w0}. 
	\item The masses $M_{uu}$, $M_{dd}$, and $M_{ss}$ are the masses of the neutral mesons $\bar{u}u, \bar{d}d$, and $\bar{s}s$, where only connected diagrams are considered in the propagators.
	They are neutral and have no magnetic moment (see \cite{BMW13}).
	\item $\Delta M^2 = (M_{dd}^2-M_{uu}^2)$ is a measure of strong-isospin-breaking that is not significantly affected by electromagnetic corrections. 
	The equality $\Delta M^2 = 2B_2\delta m$ is valid up to effects that are beyond leading order in isospin breaking, as explained in \cite{Bijnens2007}.
	\item $M^2_{\pi_\chi} = \tfrac{1}{2}(M_{uu}^2 + M_{dd}^2)$ equals the neutral pion mass $M^2_{\pi^0}$
	up to terms that are beyond leading order in isospin breaking \cite{Bijnens2007}. 
\end{itemize}

Now, the quantities $[w_0]_*$, $[M^2_{ss}]_*$ and $[\Delta M^2]_*$ are not experimentally available. 
However, they have a well-defined physical continuum limit and can thus be computed using type-I fits. 
We get the following results:
\begin{align*}
	&[w_0]_* = 0.17236(29)(63)[70] \,\text{fm}
	\\
	&[M_{ss}]_* = 689.89(28)(40)[49] \,\text{MeV}
	\\
	&[\Delta M^2]_* = 13170(320)(270)[420] \,\text{MeV}^2
\end{align*}
where the errors are statistical, systematic and total respectively.

\paragraph{Isospin decomposition} 
We write the expectation value of an observable $O$ in QCD+QED, using the set of quantities defined above (the continuum limit is assumed):
\begin{equation}
	\langle O \rangle = \langle O \rangle (M_{\pi_\chi}w_0,\,M_{ss}w_0,\,\tfrac{L}{w_0},\,\Delta M w_0,\,e),
\end{equation} 
where $\Delta M = \sqrt{\left(M_{dd}^2-M_{uu}^2\right )}$.
The QED part is defined by switching off the electric charge, keeping the other parameters at their physical values:
\begin{equation}
	\langle O \rangle_{\text{QED}} \equiv  e^2_* \cdot \tfrac{\partial \langle O \rangle }{\partial e^2} (M_{\pi_\chi}w_0,\,M_{ss}w_0,\,\tfrac{L}{w_0},\,\Delta M w_0,\,e=0).
\end{equation}
The strong isospin breaking part is defined as the differential
\begin{equation}
	\langle O \rangle_{\text{SIB}} \equiv 
	[\Delta M w_0]_*^2 \cdot \tfrac{\partial \langle O \rangle }{\partial ( \Delta M w_0)^2} (M_{\pi_\chi}w_0,\,M_{ss}w_0,\,\tfrac{L}{w_0},\,\Delta M w_0=0,\,e=0).
\end{equation}
The isospin-symmetric part is given by the remaining part, computed at $e=\Delta M w_0 = 0$,
\begin{equation}
	\langle O \rangle_{\text{ISO}} \equiv \langle O\rangle (M_{\pi_\chi}w_0, M_{ss}w_0,\,\tfrac{L}{w_0},\,\Delta M w_0=0,\,e=0).
\end{equation}

Finally, we show how $\langle O \rangle_{\text{ISO}}$, $\langle O \rangle_{\text{SIB}}$, and $\langle O \rangle_{\text{QED}}$ emerge from the fitting procedure. 
Our observable $O$ can be parametrized by the linear function 
\begin{equation}
	O = f(\{\tilde{X}\},\tilde{A},\tilde{B},...) \equiv \tilde{A} + \tilde{B}\tilde{X}_l 
	+ \tilde{C}\tilde{X}_s + \tilde{D}\tilde{X}_{\delta m} + \tilde{E}\tilde{X}_{vv} + \tilde{F}\tilde{X}_{vs} + \tilde{G}\tilde{X}_{ss},
	\label{eq:lineartype2}
\end{equation}
where the fit coefficients $\tilde{A}$, $\tilde{B}$, ... have the same form as in \eqref{eq:lineartype1}. 
If we consider separately each isospin derivative, \eqref{eq:lineartype2} can be split to a system of five equations:
\begin{equation}
	\begin{cases}
		[O]_0 = \tilde{A} + \tilde{B} \tilde{X_l} + \tilde{C} \tilde{X_s}
		\\
		[O]'_m = [\tilde{D}\tilde{X}_{\delta m}]'_m
		\\
		[O]''_{20} = [\tilde{A}+\tilde{B}\tilde{X}_l+\tilde{C}\tilde{X}_s+\tilde{D}\tilde{X}_{\delta m}]''_{20} + [\tilde{E}]_0 
		\\
		[O]''_{11} = [\tilde{A}+\tilde{B}\tilde{X}_l+\tilde{C}\tilde{X}_s+\tilde{D}\tilde{X}_{\delta m}]''_{11} + [\tilde{F}]_0 
		\\
		[O]''_{02} = [\tilde{A}+\tilde{B}\tilde{X}_l+\tilde{C}\tilde{X}_s+\tilde{D}\tilde{X}_{\delta m}]''_{02} + [\tilde{G}]_0 
	\end{cases}
\end{equation}
We get the isospin-breaking decomposition after a continuum extrapolation is performed:
\begin{equation}
	\langle O\rangle_{\mathrm{ISO}} = \tilde{A}_0,
	\qquad
	\langle O\rangle_{\mathrm{SIB}} = \tilde{D}_0\tilde{X}_{\delta m},
	\qquad
	\langle O\rangle_{\mathrm{QED}} = e_*^2 (\tilde{E}_0+\tilde{F}_0+\tilde{G}_0),
\end{equation}
where $\langle O\rangle_{\mathrm{QED}} = e_*^2 (\tilde{E}_0+\tilde{F}_0+\tilde{G}_0)$ can be further separated in 
\begin{equation}
	\langle O\rangle_{\mathrm{QED-vv}} = e_*^2 \tilde{E}_0, 
	\qquad
	\langle O\rangle_{\mathrm{QED-vs}} = e_*^2 \tilde{F}_0, 
	\qquad
	\langle O\rangle_{\mathrm{QED-ss}} = e_*^2 \tilde{G}_0. 
\end{equation}

\section{Conclusions}
QED and strong-isospin-breaking corrections to $a_\mu^{\mathrm{LO-HVP}}$ are necessary to reach the precision needed for comparison to experiments.
In our work, we have included IB corrections, in current propagators and in hadron masses, by expanding these quantities to first order in the isospin-breaking parameters $\delta m$ and $e^2$, and by measuring each term on isospin-symmetric configurations.
Furthermore, we have defined an isospin decomposition of observables, which allows us to evaluate the QED and the strong-isospin-breaking contributions to $a_\mu^{\mathrm{LO-HVP}}$ summarized in {Table \ref{tab:our_result}}. 

A detailed comparison of isospin-breaking effects computed by different collaborations is a delicate matter, as contributions computed and separation scheme used may differ from group to group.
It is nevertheless essential to mention the other works on the subject.
Connected valence-valence QED effects have been computed in \cite{Blum2018} and \cite{Giusti2019}.
In particular, the strange contribution to this effect is reported in  \cite{Giusti2017} and in the supplemental material of \cite{Blum2018}.
The connected part of the strong-isospin-breaking effect has been computed in \cite{Giusti2019, Blum2018, Chakraborty2018} and most recently in \cite{Lehner2020}.
See also \cite{gerardin_anomalous_2021} for a comparative table and further details.

\begin{table}
	\begin{center}
		\begin{tabular}{l|c|c|c}
		& $a_\mu^{\mathrm{strange}}(L_{\mathrm{ref}},T_{\mathrm{ref}})$ & $a_\mu^{\mathrm{light}}(L_{\mathrm{ref}},T_{\mathrm{ref}})$ & $a_\mu^{\mathrm{disc}}(L_{\mathrm{ref}},T_{\mathrm{ref}})$
		\\ \hline\hline
		QED-vv & {-0.0086(42)(41)} & {-1.24(40)(31)} & {-0.55(15)(10)}
		\\ \hline
		QED-vs & { -0.0014(11)(14)} & { -0.0079(86)(94)} & { 0.011(24)(14)}
		\\ \hline
		QED-ss & { -0.0031(76)(69)} & { 0.37(21)(24)} & { -0.040(33)(21)}
		\\ \hline
		{Total QED} & -0.0136(86)(76) & -0.93(35)(47) & -0.58(14)(10) 
		\\ \hline	
		{SIB} & 0 & {6.60(63)(53)} & {-4.67(54)(69)}
		\\ \hline\hline	
		{Total IB} &  -0.0136(86)(76) & 5.67(72)(71) & -5.25(56)(70)
		\\ \hline
		\end{tabular}
	\end{center}
	\caption{Continuum extrapolated results for the different components of the strange, light and disconnected,
	strong and QED isospin-breaking contributions to $a_\mu^{\mathrm{LO-HVP}} \times 10^{10}$. 
	Labels \textit{v} and \textit{s} refer to \textit{valence} and \textit{sea} respectively. 
	The first error is statistical, the second systematic. 
	These results correspond to a box of size $L_{\mathrm{ref}} = 6.27~\mathrm{fm}$ and $T_{\mathrm{ref}} = \tfrac{3}{2} L_{\mathrm{ref}}$.}
	\label{tab:our_result}
\end{table}

\end{document}